\documentclass[
    ,final            
  ]
  {aipproc}

\layoutstyle{6x9}

\begin{document}

\title{General Relativistic MHD Simulations of the Gravitational Collapse of a 
Rotating Star with Magnetic Field as a Model of Gamma-Ray Bursts}

\author{Y. Mizuno}{
  address={Department of Astronomy, Kyoto University, Kyoto 606-8502}
}

\author{S. Yamada}{
  address={Department of Science and Engineering, Waseda University, Tokyo 169-8555}
}

\author{S. Koide}{
  address={Department of Engineering, Toyama University, Toyama 930-8555}
}

\author{K.Shibata}{
  address={Kwasan and Hida Observatory, Kyoto University, Kyoto 607-8471}
}

\begin{abstract}
 We have performed 2.5-dimensional general relativistic magnetohydrodynamic (MHD) simulations of the gravitational collapse of a magnetized rotating massive star as a model of gamma ray bursts (GRBs). This simulation showed the formation of a disk-like structure and the generation of a jet-like outflow inside the shock wave launched at the core bounce. We have found the jet is accelerated by the magnetic pressure and the centrifugal force and is collimated by the pinching force of the toroidal magnetic field amplified by the rotation and the effect of geometry of the poloidal magnetic field. The maximum velocity of the jet is mildly relativistic ($\sim$ 0.3 c).
\end{abstract}

\maketitle


\section{Introduction}

GRBs and the afterglows are well described by the fireball model, in which a relativistic outflow is generated from a compact central engine. Rapid temporal decay of several afterglows is consistent with the evolution of a highly relativistic jet with bulk Lorentz factors $\sim 10^{2} - 10^{3}$. The formation of relativistic jets from a compact central engine remains one of the major unsolved problems in GRB models. 

From recent observations, some evidence was found for a connection between GRBs and the death of massive stars. This evidence includes a correlation between star forming regions and the position of GRBs inside the host galaxy \citep{Blo02a}, a \lq\lq bump" resembling the light curves of Type Ic supernovae in the optical afterglow of several GRBs \citep{Blo02b, Gra02}, and association of GRB980425-SN1998bw \citep{Iwa98, Woo99} and GRB030329-SN2003df \citep{Sta03, Hjo03}. Several authors \citep{Fra01, Pan01, Blo03} have studied beaming angles and energies of a number of GRBs. They have found that central engines of GRBs release supernova-like energies ($\sim 10^{51}$ erg). It is thus probable that a major subclass of GRBs is a consequence of the collapse of a massive star. 

In this study, we perform 2.5-dimensional general relativistic MHD simulations of the gravitational collapse of a rotating star with magnetic field as a model for a collapsar. The collapsar is in some sense an anisotropic supernova, and it is considered that relativistic jets from collapsars are launched by MHD processes in accreting matter and/or by neutrino annihilation \citep{Woo93, Mac99}. 

\section{Numerical Method}

In order to study the formation of relativistic jets from a collapsar we use a 2.5-dimensional general relativistic magnetohydrodynamics (GRMHD) code \citep{Koi03} .We assume that the gravitational potential is constant in time. We neglect nuclear burnings and neutrinos in this simulation.

We consider the following situation as the initial condition for the simulations: A few $M_{\odot}$ black hole is produced at the center of the stellar remnant with a weak shock standing at a radius of a few hundred km and the post-shock gas falling onto the central black hole.
We consider a non-rotating black hole as the central black hole. We employ 1-dimensional supernova simulation data of Bruenn (1992) \citep{Bru92} to obtain the initial density, pressure and radial velocity distribution. We add the effect of stellar rotation and intrinsic magnetic field. The initial rotational velocity distribution is assumed to be a function of the distance from the rotation axis only. The initial magnetic field is assumed to be uniform and parallel to the rotational axis. See Mizuno et al. (2003) \citep{Miz03} for details. 

\section{Results}
\subsection{Formation of Jet}

\begin{figure}
  \includegraphics[height=.5\textheight]{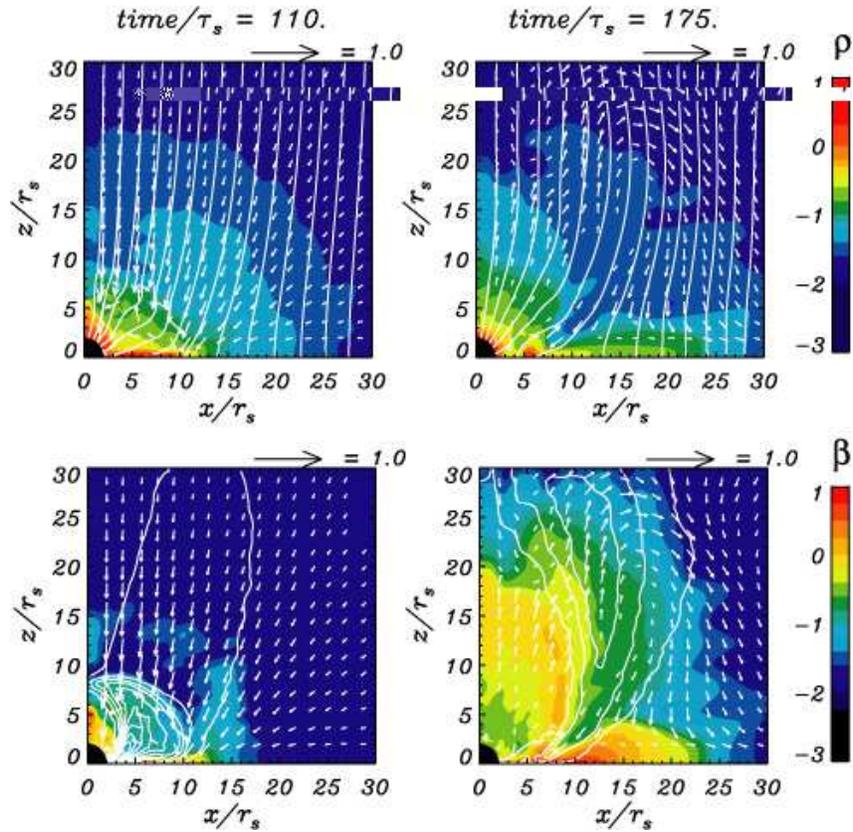}
  \caption{The time evolution of the density (upper panel) and the plasma beta (lower panel). The color scale shows the value of the logarithm of density and the plasma beta. The white curves depict magnetic field lines (upper panel) and contour of the toroidal magnetic field (lower panel) . Arrow represent the poloidal velocities normalized by the light velocity. Accreting matter forms a disk-like structure. The shock wave is generated near the central black bole. The jet-like outflow is produced inside the shock wave.}
\end{figure}

The stellar matter falls onto the central black hole at first. This collapse is anisotropic due to the effects of rotation and magnetic field. The accreting matter falls more slowly on the equatorial plane than on the rotational axis. The matter piles up on the equatorial plane, and a disk-like structure is formed near the central black hole. Since the magnetic field is frozen into the plasma, it is dragged by the accreting matter and amplified. The amplified magnetic field expands outwards as Alfv\'{e}n waves and launches an outgoing shock wave. 

The jet-like outflow is generated behind the shock wave. The shock wave has the high magnetic pressure gradient force and high centrifugal force. These forces push back the accreting matter and construct the jet-like outflow. The jet-like outflow formed in the simulation is thus magnetically driven. The jet has a mildly relativistic velocity, $ \sim 0.3$ c (the poloidal velocity is $ \sim 0.1$ c). It exceeds the escape velocity. Thus, the jet is likely to get out of the stellar remnant. The magnetic field plays an important role in the collimation. Not only the pinching force of the toroidal magnetic field but also the geometry of the poloidal magnetic field (i.e. poloidal magnetic pressure) plays a crucial role.

\subsection{Dependence on the Initial Magnetic Field Strength and Initial Rotatinal Velocity}

As the initial magnetic field strength increases, the jet velocity increases and the magnetic twist decreases. However, for stronger magnetic field the jet velocity decreases with increasing initial magnetic field strength and the magnetic twist still continues decreasing.  
As the initial rotational velocity becomes faster, the jet velocity becomes faster and the magnetic twist becomes stronger up to a certain value. For faster initial rotational velocity, the vertical component of the jet velocity and the magnetic twist are almost constant.

In order to produce a strong jet, the magnetic field has to be twisted significantly so that it can store enough magnetic energy. If the initial magnetic field is strong, the magnetic field cannot be twisted significantly because Alfv\'{e}n waves propagate as soon as the magnetic field is twisted a little bit. As a result, the jet velocity does not rise up and the magnetic twist remains weak. Therefore, weaker initial magnetic fields are favorable for a stronger jet. The dependence on the initial rotational velocity can be understood by a same reason as the dependence on the initial magnetic field strength. If the initial rotation is sufficiently fast, the magnetic field is twisted significantly and stores enough energy to produce a fast jet. However, the stored magnetic energy has a limit. It is determined by the competition between the propagation time of Alfv\'{e}n waves and the rotation time of the disk (or the twisting time).

\section{Discussion}

We have studied the generation of a jet from gravitational collapse of a rotating star with magnetic field by using the 2.5D general relativistic MHD simulation code. The maximum velocity of the jet is mildly relativistic ($\sim$ 0.3c). This result is consistent with Newtonian case \citep{Pro03}. The jet is too slow for the jet of GRBs. We have to consider other acceleration mechanism. The break out of the jet through the stellar surface is the most applicable acceleration mechanism. When the jet goes through the stellar surface, the strong density gradient may accelerate the jet. In fact, some authors \citep{Alo00, Zha03} have showed numerically that a significant acceleration of the jet occur and the terminal Lorentz factor becomes as high as $\Gamma \sim 50$. We think it is important to simulate the propagation of the jet outside the stellar surface by a GRMHD code properly evaluating the importance of magnetic fields for the dynamics and further propagation of the jet.


\begin{theacknowledgments}
  This work was partially supported by a Grant-in-Aid for the 21st Century COE ``Center for Diversity and Universality in Physics'' and Grants-in-Aid for the scientific resarch from Ministry of Education, Science, Sports, Technology and Culture of Japan through No.14079202, No.14540226, and No.14740166.
  
\end{theacknowledgments}

\end{document}